\documentclass[%
 reprint,
superscriptaddress,
nobibnotes,
 amsmath,amssymb,
 aps, prl
]{revtex4-2}

\usepackage{graphicx}
\usepackage{dcolumn}
\usepackage{bm}
\usepackage{hyperref}
\usepackage{float}
\usepackage{siunitx}
\usepackage{gensymb}
\usepackage{xcolor}

\usepackage{soul,xcolor}
\setstcolor{red}

\newcommand{\drond}[2]{\frac{\partial #1}{\partial #2}}
\newcommand{\drondcarre}[2]{\frac{\partial^2 #1}{\partial #2^2}}

\newcommand{\iemn}{Université de Lille, CNRS, Centrale Lille, Université Polytechnique Hauts-de-France, UMR 8520 - IEMN - Institut d'Electronique de Microélectronique et de Nanotechnologie, F-59000 Lille, France}

\begin{document}

\preprint{APS/123-QED}

\title{Oscillations and Cavity Modes in the Circular Hydraulic Jump}

\author{Aurélien Goerlinger}
\email{aurelien.goerlinger@univ-lille.fr}
\affiliation{\iemn}
\author{Michael Baudoin}
\affiliation{\iemn}
\affiliation{Institut Universitaire de France, 1 rue Descartes, 75005 Paris}
\author{Farzam Zoueshtiagh}
\author{Alexis Duchesne}
\affiliation{\iemn}

\date{\today}

\begin{abstract}
We report spontaneous oscillations of circular hydraulic jumps created by the impact of a submillimeter water jet on a solid disk. The jet flow rate is shown to condition the occurrence of the oscillations while their period is independent of this parameter. The period, however, varies linearly with the disk radius. This dependency is rationalized by investigating surface gravity wave modes in the cavity formed by the disk. We show that the jump oscillation frequency systematically matches one of the surface wave disk-cavity eigenfrequencies, leading to the conclusion that the oscillations are self-induced by the interaction between the jump and surface wave eigenmodes.
\end{abstract}

\maketitle



\begin{figure}
    \centering
    \includegraphics[width=.9\linewidth]{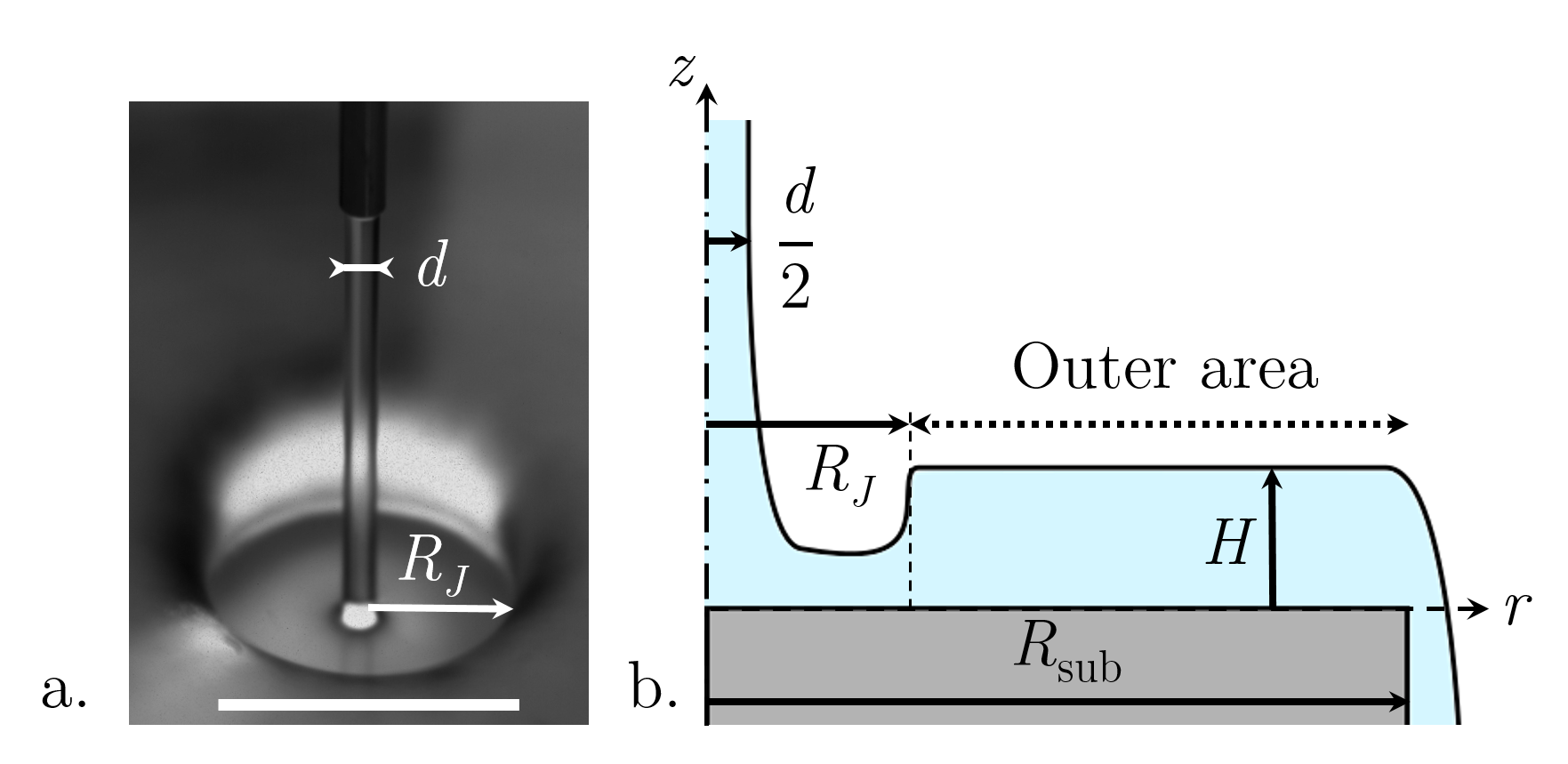}
    \includegraphics[width=.9\linewidth]{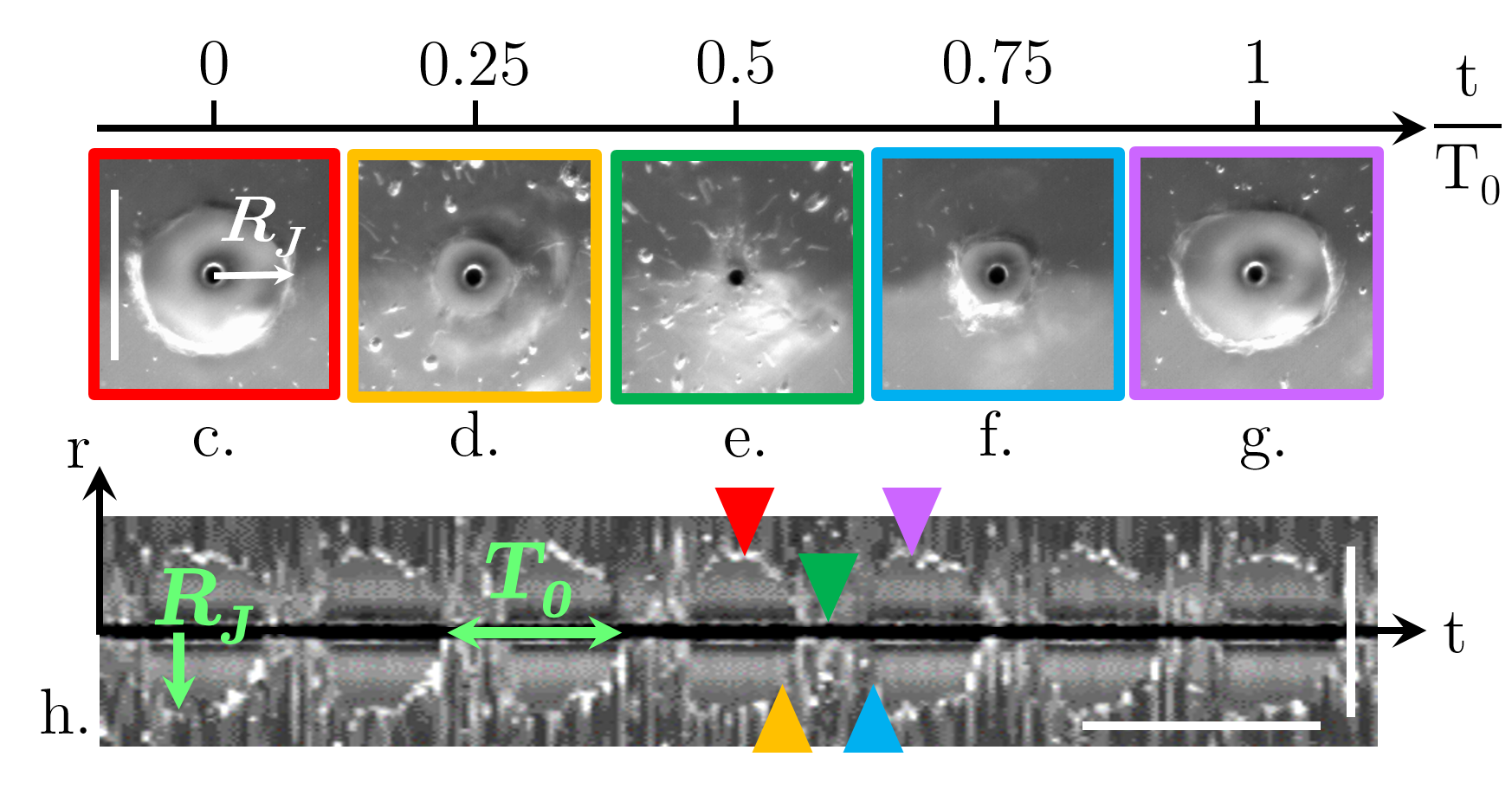}
    \caption{\textbf{(a)} Photography of a circular hydraulic jump. The water jet diameter $d$ is 0.9 mm. White scale bar is 1 cm. \textbf{(b)} Sketch of the height profile of a circular hydraulic jump. \textbf{(c) to (g)} Images of an oscillating circular hydraulic jump taken from below with a fast camera at five different times. $T_0=0.37$~s corresponds to the period of oscillation. White scale bar is 1 cm. \textbf{(h)} Spatio-temporal evolution of the diameter of the oscillating circular hydraulic jump. Vertical white scale bar is 1 cm, horizontal white scale bar is 0.5 s. See movie M1 in Supplemental Material \cite{SuppMat} for an animated representation of the jet oscillation.}
    \label{fig:oscillation}
\end{figure}

The hydraulic jump is a ubiquitous phenomenon occurring when a liquid jet impinges on the bottom of a sink. It is typically characterized by a circular thin liquid film around the impact point of the jet, which suddenly and substantially thickens at a certain radial distance (Fig. \ref{fig:oscillation} (a)). Hydraulic jumps can also occur naturally on channels, rivers, downstream of a dam or in the case of tidal bores going upstream \cite{chansonCurrentKnowledgeHydraulic2009}. 
These flow discontinuities separate an inner flow that is thin and supercritical (faster than surface gravity waves) from an outer thicker subcritical flow. The circular hydraulic jump has been the subject of extensive experimental, numerical and theoretical investigations since the pioneering works of Rayleigh \cite{rayleighTheoryLongWaves1914} and Tani \cite{taniWaterJumpBoundary1949}. Many efforts have been made to predict the radius of the jump in the steady state, with notable theories proposed by Watson \cite{watsonRadialSpreadLiquid1964} and Bohr \cite{bohrShallowwaterApproachCircular1993} which have been further discussed and improved in subsequent works \cite{liuHydraulicJumpCircular1993,bushInfluenceSurfaceTension2003,duchesneConstantFroudeNumber2014,mohajerCircularHydraulicJump2015,chooInfluenceNozzleDiameter2016,rojasProgressiveCorrectionCircular2013}. However, the influence of surface tension on the radius of the jump remains a highly debated topic \cite{bhagatOriginCircularHydraulic2018, duchesneSurfaceTensionOrigin2019,bohrSurfaceTensionEnergy2021,bhagatCircularHydraulicJump2022,duchesneCircularHydraulicJumps2022,wangEffectsGravitySurface2021}. In addition to the stable stationary cases, it has been discovered that the circular hydraulic jump can become unstable under certain conditions and transition to polygonal or even more complex shapes \cite{ellegaardCoverIllustrationPolygonal1999,bushExperimentalInvestigationStability2006,martensModelPolygonalHydraulic2012,laboussePolygonalInstabilitiesInterfacial2015}, and Wang \cite{wangInfluenceAzimuthallyVarying2022} observed noncircular shapes for disks with azimuthally varying edge conditions. Furthermore, non stationary cases have also been reported. Liu \cite{liuHydraulicJumpCircular1993} and Rao \cite{raoWaveStructureRadial2001} observed a transition to turbulence, while Teymourtash \cite{teymourtashExperimentalInvestigationStationary2015} observed rotating jumps. Finally, Craik \cite{craikCircularHydraulicJump1981} investigated the behavior of the circular hydraulic jump as the depth of the outer area is increased and found a transient oscillatory instability of the jump radius. However, to the best of our knowledge, stable periodic states have not been described in the literature yet.

Here, we report on the stable periodic oscillations of hydraulic jumps (see Fig. \ref{fig:oscillation} and Supplemental Material \cite{SuppMat} M1) produced by a submillimetric water jet impinging a Plexiglas disk. First, we explore the parameters (flow rate, disk radius) leading to this instability. Second, we demonstrate that the oscillation periods systematically match those of the surface gravity wave modes of the disk-cavity system. Finally, we verify one of the predictions of our model by showing two synchronous oscillating jumps in phase opposition on the same disk. \\

The hydraulic jump is formed in the experiments by a jet of deionized water flowing through a needle with an inner diameter of 0.84~mm (according to the manufacturer, Weller) and impinging on a right angle-edged Plexiglas disk placed 1~cm below. The Plexiglas disk is 1~cm thick and has different radii, $R_\text{sub}$, ranging from 1 to 6~cm. The effective water jet diameter was measured prior to experiments and found to have a diameter of 0.9~mm. The flow rate $Q$ (2 to 3 \si{\milli\litre\per\second}) is controlled by adjusting the driving pressure with a constant-level water reservoir whose height ranges from 50~cm to 2~m above the disk plane. The use of this constant-level reservoir ensures a constant flow rate during each experiment, and we verified the jet flow was stable by zooming on it with a camera at 60~FPS. The Weber number ranges from 130 to 290 while the Reynolds number ranges from 2800 to 4250. The hydraulic jump dynamics are observed from below through the transparent Plexiglas disk with a 45\degree \ angled mirror, which directs the image into a Mikrotron Motionblitz Cube 4 camera equipped with a Tamron SP AF28-75mm camera lens, acquiring images at 46 FPS. Videos are processed with ImageJ software and a Python code.

\begin{figure}
    \centering
    \includegraphics[width=.45\textwidth]{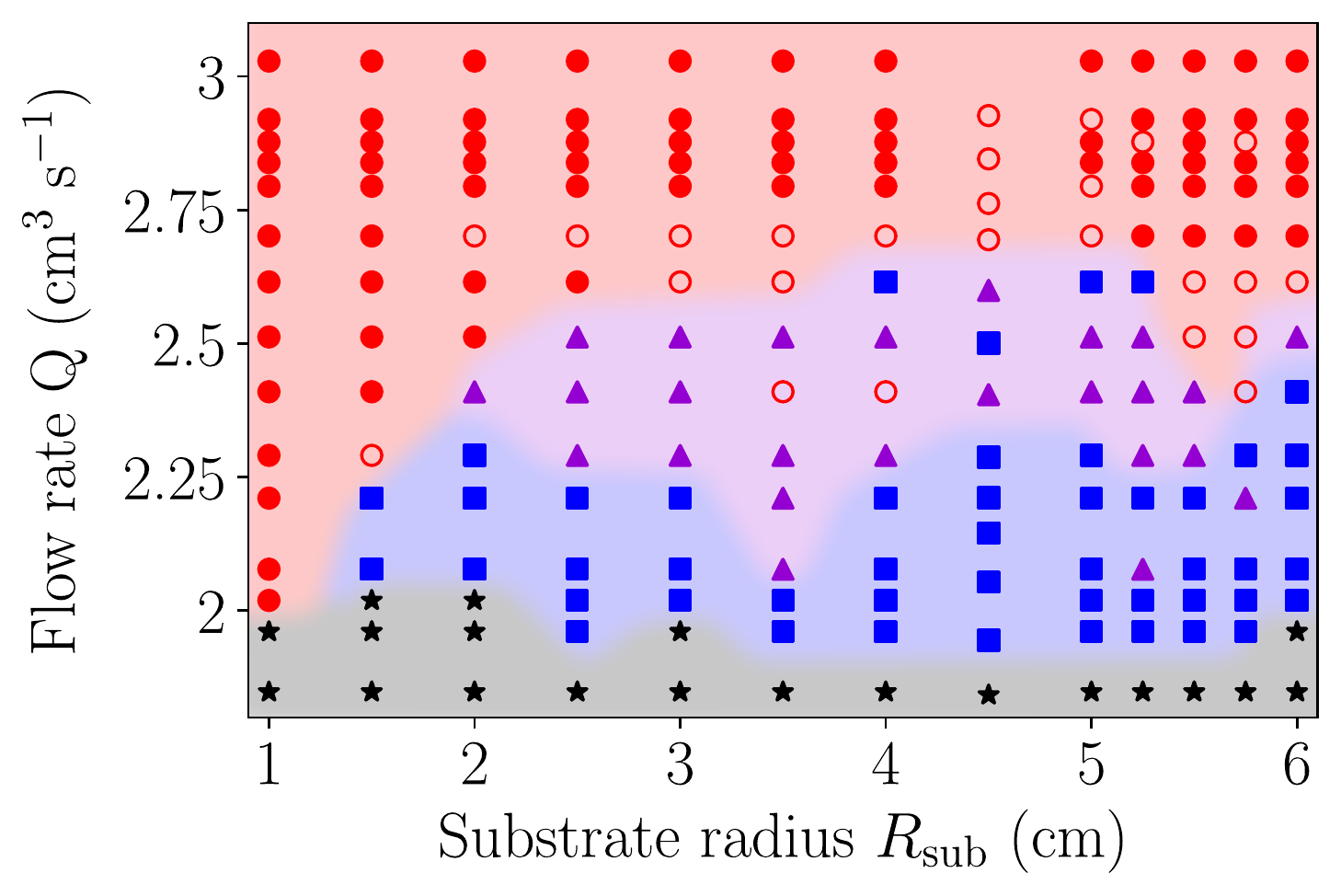}
    \caption{Phase diagram illustrating the behavior of the circular hydraulic jump with respect to the flow rate $Q$ and disk radius $R_\text{sub}$. The water jet impinges on the disk at its center. Red points represent stationary jumps, red circles correspond to unstable oscillating jumps which quickly (i.e., under a minute at most) come back to a stationary state, and purple triangles represent a bistable state in which the jump can either oscillate periodically or stay stationary. Finally, blue squares represent stable oscillating jumps and black stars correspond to situations wherein the jump cannot stay open. Each point corresponds to at least 3 experiments.}
    \label{fig:diagramme}
\end{figure}

The dynamics observed for different flow rates ranging from 2~\si{\milli\litre\per\second} to 3~\si{\milli\litre\per\second} and disk radii ranging from 1~cm to 6~cm are summarized on the phase diagram shown in Fig.  \ref{fig:diagramme}. At the highest flow rates (around 3~\si{\milli\litre\per\second}), stationary jumps with a constant radius are observed for any disk radius (indicated by red dots in Fig. \ref{fig:diagramme}). As the flow rate is decreased, a transient state is observed (represented by red circles) where the jump radius oscillates between zero (no jump) and a maximum value for a few dozens of seconds before finally reaching a stationary state. For flow rates around 2.5~\si{\milli\litre\per\second}  and disk radii larger than 2~cm, a bistable state is observed (purple triangles). In this state, the system exhibits two stable behaviors for the same driving parameters: either stable periodic oscillations or a stationary jump. However, as the flow rate decreases below approximately 2.2~\si{\milli\litre\per\second}, the bistable state disappears, and the system exhibits systematic stable periodically oscillating jumps (indicated by blue squares). Figures \ref{fig:oscillation}(c) to \ref{fig:oscillation}(h) show an example of such stable periodic jump oscillations. Finally, for flow rates below 1.8~\si{\milli\litre\per\second}, the hydraulic jump ceases to occur as it is unable to open itself more than a few times at the start of the experiment before remaining indefinitely closed (represented by black stars in Fig. \ref{fig:diagramme}). It is interesting to note that the stable oscillatory state is only observed within a narrow range of parameters.

\begin{figure}
    \centering
    \includegraphics[width=.45\textwidth]{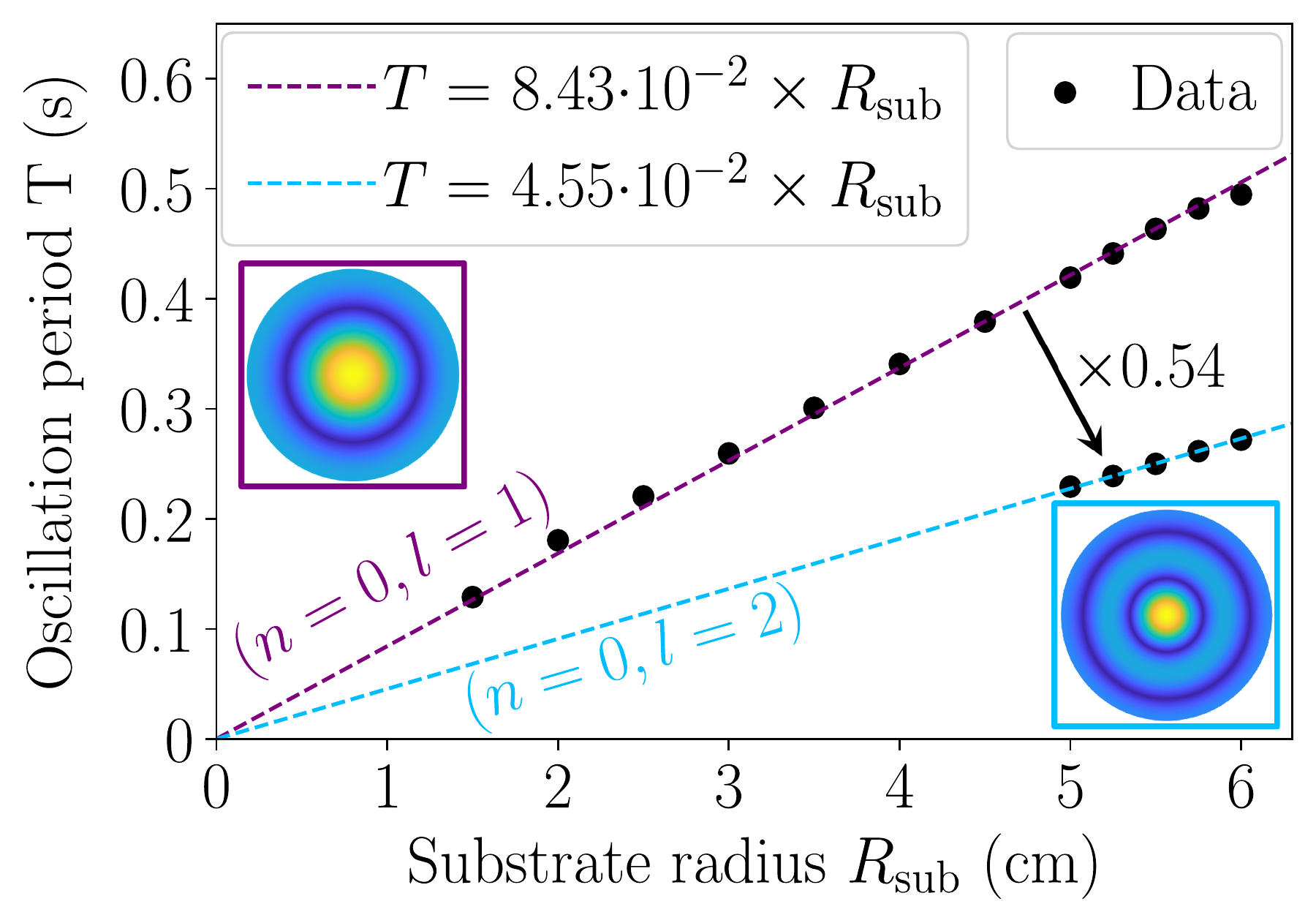}
    \caption{Evolution of the oscillation period $T$ (averaged over 50 periods) with respect to the disk radius $R_\text{sub}$. The purple dashed line corresponds to the linear fit applied to the fundamental periods while the blue dashed line represents to the linear fit applied to the harmonic periods. Both insets show the relative amplitude of the 0th order Bessel function of the first kind $J_0$ in polar coordinates (small amplitudes correspond to deep blue area while high amplitudes correspond to yellow areas). The purple-edged inset illustrates the fundamental mode where the second antinode is at the disk's border while the blue-edged inset pictures the harmonic mode where the third antinode is at the disk's border. Both modes are respectively characterized with 2 integers $n$ and $l$.}
    \label{fig:periode}
\end{figure}

We further investigated the dependency of the oscillations on the experimental parameters $Q$ and $R_\text{sub}$. The data analysis revealed that the period of oscillation, measured with an accuracy of 0.3~ms, does not depend on $Q$ even though $Q$ conditions the emergence of the oscillations. However, the period does show a linear dependence on the disk radius, as depicted in Fig. \ref{fig:periode}. Notably, for disk radii greater than $R_\text{sub} = 5$ cm, the data points exhibit two distinct linear trends with different slopes. This implies that, in the same experimental conditions, the observed oscillation period is one of these two values. In the following, we refer to the modes associated with the larger and lower slopes as the ”fundamental” (in purple in Fig. \ref{fig:periode}) modes and ”harmonic” (in blue) modes, respectively. Interestingly the ratio between these two slopes is found to be equal to 0.54.

To rationalize these observations, we investigated the possibility that surface gravity wave modes inside the cavity formed by the thick outer subcritical region could be the origin of this instability, as it is known that hydraulic jumps emit surface gravity waves in this area \cite{hansenGeometricOrbitsSurface1997,rayStandingTravellingWaves2007}. Since the velocity of the surface wave depends on the height of the fluid layer, the brutal flow variations at the edge of the disk act as a barrier, i.e., a reflector for these waves. The outer layer allows the upstream propagation of these surface waves, and when they are reflected at the edge of the disk, they can travel back toward the jump, giving rise to standing waves. A theoretical model of the resonance modes of the disk cavity can be derived from Saint-Venant's equations, which describes the deformations of a free surface under shallow water conditions and can be written as \cite{saint-venantTheorieMouvementNon1871}
\begin{align*}
    \drond{h}{t}+\drond{}{x}\left((H+h)u\right)+\drond{}{y}&\left((H+h)v\right) = 0 \ ;\\
    \drond{u}{t}+u\drond{u}{x}+v\drond{u}{y}-fv=-g\drond{h}{x}-&ku+\nu\left(\drondcarre{u}{x}+\drondcarre{u}{y}\right);\\
    \drond{v}{t}+u\drond{v}{x}+v\drond{v}{y}-fv =-g\drond{h}{y}-&kv+\nu\left(\drondcarre{v}{x}+\drondcarre{v}{y}\right).
\end{align*}
Here, $u$ and $v$ represent $x$ and $y$ components of the flow velocity, $H$ is the average height of the free surface, $h$ is the deformation of the free surface from $H$, $f$ is the Coriolis coefficient, $k$ is the viscous drag coefficient and $\nu$ is the cinematic viscosity.
Under the assumptions that the deformations of the free surface are small compared to the thickness of the water layer ($h\ll H$) and that the Coriolis force and viscous dissipation can be neglected, the linearization of the Saint-Venant equations leads to the classical surface gravity waves equation in shallow water:
\begin{align*}
\frac{\partial^2 h}{\partial t^2}-c_s\Delta_s h = 0,
\end{align*}
where $c_s=\sqrt{gH}$ represents the gravity wave speed and $\Delta_s=\frac{\partial^2}{\partial x^2}+\frac{\partial^2}{\partial y^2}$ is the 2D Laplacian operator. 
The surface waves modes in the disk cavity can be obtained from the separate variable harmonic solutions of this equation in polar coordinates $(r,\theta)$ written as $h(r,\theta,t) = \sum_{n=-\infty}^{\infty} \left[ E_n J_n(k r) + F_n Y_n (k r) \right] e^{i (n \theta + \omega_n t)}$ where $E_n$ and $F_n$ are constants, $J_n$ and $Y_n$ are the Bessel functions of the first and second kind, respectively, $\omega_n = 2 \pi / T_n$ is the angular eigenfrequency of mode $n$, $T_n$ is its period and $k_n = \omega_n / c_s$ is its wave number. However, the radius of the jump is much smaller than the wavelength (i.e., $kR_J \ll 1$) and thus the jump cannot resolve the singularity of the $Y_n$ functions when $r \rightarrow 0$. They must hence be eliminated, leading to $F_n = 0$ $\forall n$. Since the water jet impinges the disk on its center, the mode must be axisymmetric with an antinode at its center, which is only fulfilled by the mode $n=0$ (associated with the Bessel function $J_0$). Additionally, since the surface wave can freely oscillate on the disk border, $r = R_{sub}$ corresponds to a surface wave antinode. 
For the fundamental mode, the disk border corresponds to the first antinode ($l=1$), which is the first maximum of the Bessel function $J_0$ (for $r>0$). This yields $k_0^1 R_{sub} = 2 \pi R_{sub} / T_0^1 c_s = 3.83$, with $k_n^l$ and $T_n^l$  the wave number and period of the $(n,l)$ mode, respectively, and $l$ the number of the Bessel function antinode at the disk border. 
For the harmonic mode, the disk border will correspond to the second antinode ($l=2$) and thus $k_0^2 R_{sub} = 2 \pi R / T_0^2 c_s = 7.02$. Based on these theoretical considerations, we obtain $T_0^2 / T_0^1 = 0.54$, which is in perfect agreement with the experimentally determined value. The two modes $(n,l) = (0,1)$ and $(n,l) = (0,2)$  are illustrated in the surface plots insets of Fig.  \ref{fig:periode}. Moreover, the model predicts a linear correlation between the period of the oscillations and the size of the cavity (i.e., the radius of the disk). This prediction aligns once again very well with our experimental observations. Finally, it is worth noting that this model strongly assumes the height profile $H$ of the water layer in the outer region of the jump, and hence the surface gravity wave speed $c_s$, to be independent of both the flow rate and the disk radius.
Besides, matching the model with the data presented in Fig. \ref{fig:periode} yields a predicted value of $3.9$ mm for $H$. To investigate the dependence of $H$ in the outer region on $Q$ and $R_\text{sub}$, we conducted experiments for 3 different disk radii -- 2.5, 4, and 5.5 cm -- as well as for 4 values of flow rate ranging from 2 \si{\milli\litre\per\second} to 3 \si{\milli\litre\per\second}. In the case of oscillating jumps, we observed periodic variations in the height profile. To compare our results for all parameters, we averaged the measured values of $H$ over time. As predicted, the height profile was found to be independent of both $Q$ and $R_\text{sub}$. Additionally, the measured mean value of $H$ was determined to be $H = 3.98 \pm 0.33$ mm, which aligns remarkably well with the prediction of the model. Therefore, the model accurately reproduces the trends depicted in Fig. \ref{fig:periode}, strongly indicating that the jump oscillations originate from an interaction with surface gravity wave modes.

In order to further challenge this description, we conducted additional investigations to explore the generation of the higher order modes $(n>0)$ predicted by the theory. Since the $n=0$ mode is the only axisymmetric mode with an antinode at the center, these higher-order modes can only be excited by positioning the jet away from the center. Hence, we systematically shifted the position of the jet along the radius of the disk in increments of 10\% of $R_\text{sub}$. For each position, we measured the period of the observed oscillations. This process was repeated for different disk radii ranging from 3 to 5 cm. 

\begin{figure}
    \centering
    \includegraphics[width=.45\textwidth]{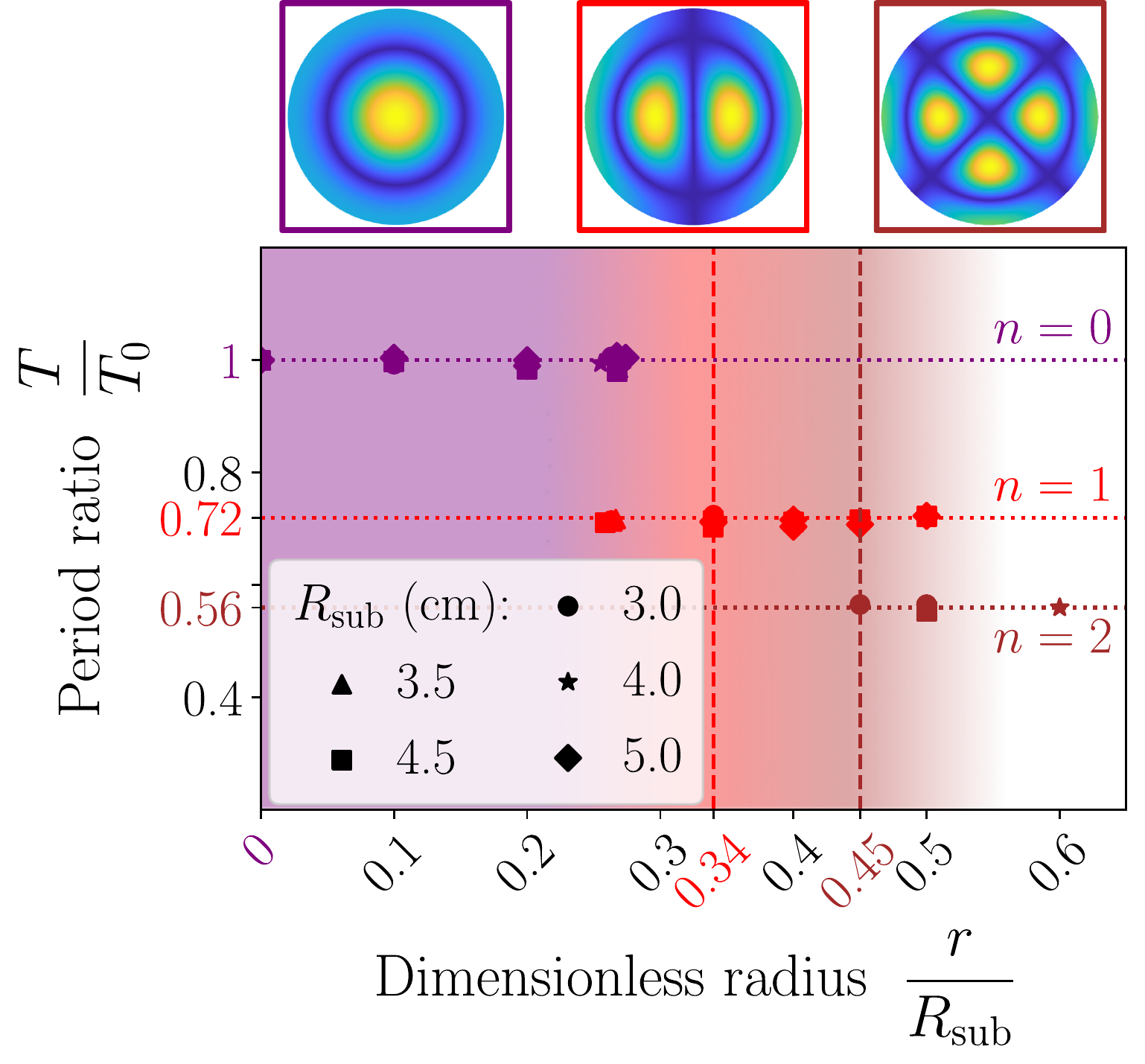}
    \caption{Ratio of the observed oscillation period $T$ and the period $T_0$ of the $(n,l) = (0,1)$ mode with respect to the ratio between the distance $r$ of the jet center to the center of the disk and the disk radius $R_\text{sub}$, for different disk radii. The horizontal dotted lines show the predicted period ratio $\frac{T}{T_0}$ for the $n=0$ (purple), $n=1$ (red) and $n=2$ (brown) modes while the vertical dashed lines show the relative positions of the first antinode of mode $n=1$ and $n=2$. The vertical dashed lines show the position of the closest antinode to the center for all three modes. Color edged insets show the relative amplitude of the Bessel function associated to all three modes.}
    \label{fig:higher_modes}
\end{figure}

Figure \ref{fig:higher_modes} shows the period of measured oscillations as function of the relative position of the jet with respect to the center. When the jet impinges on the disk close to its center, we measured the period $T_0$ associated to the $(n,l)=(0,1)$ mode. As the the distance between the jet and the center increases and approaches approximately $0.29 \ R_\text{sub}$, the oscillation period decreases by a factor of 0.72 compared to $T_0$. Furthermore, as the distance increase beyond $0.5 \ R_\text{sub}$ the oscillation period decreases even further by a factor of 0.56 relative to $T_0$.
However, no further changes in the period are observed beyond this point, and the hydraulic jump ceases to oscillate for distances $r>0.6 \, R_\text{sub}$. Throughout these experiments, no influence of the disk radius was observed whatsoever.

\begin{figure}
    \centering
    \includegraphics[width=.45\textwidth]{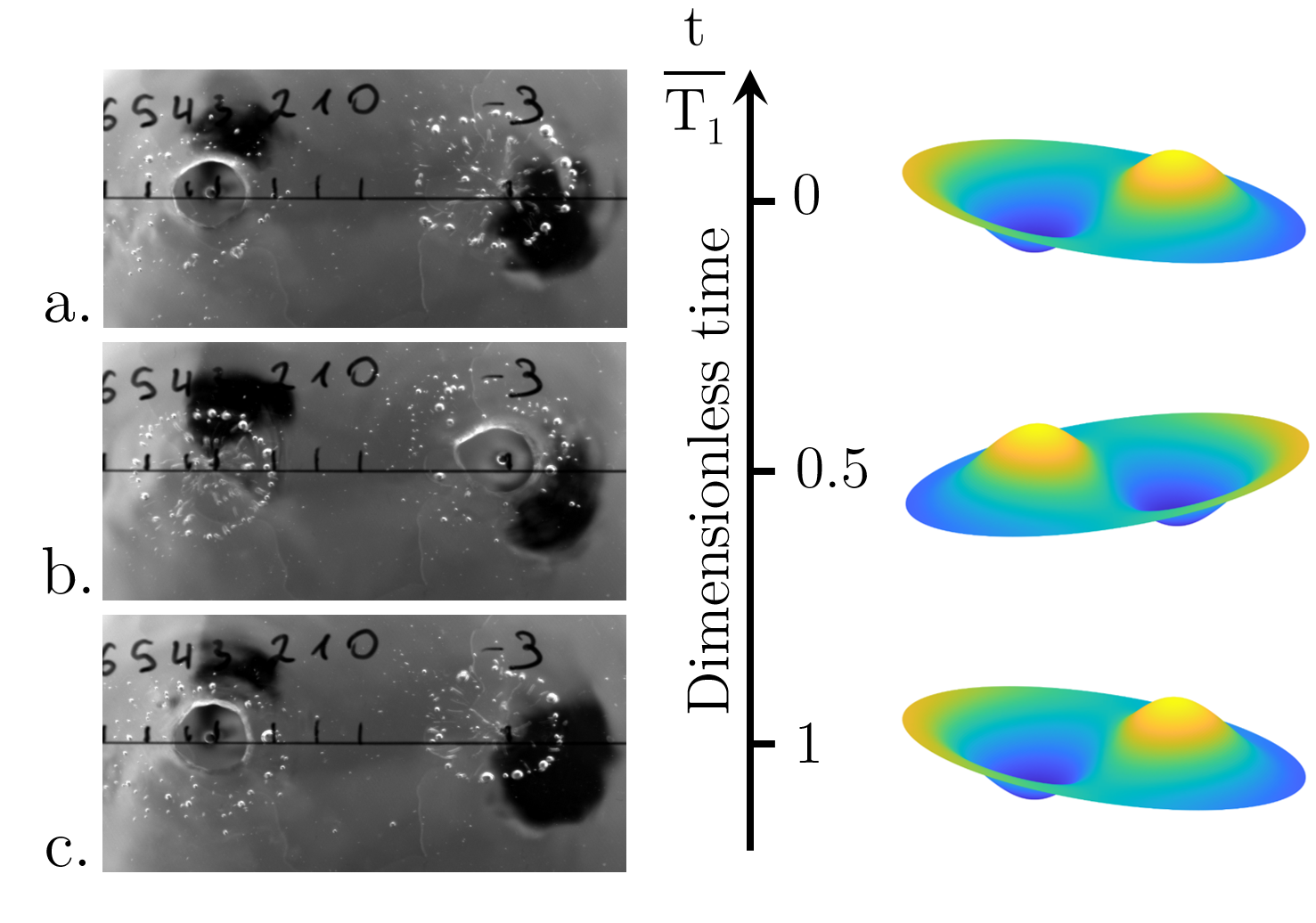}
    \caption{Two hydraulic jumps oscillating in phase opposition in a dual jet experiment performed  on  a disk of 4.5 cm radius. Both jets ('-3' and '3' graduations) are at a relative distance of $0.34 \, R_\text{sub} = 1.53$ cm from the disk center ('0' graduation). \textbf{(a) to (c)} Left: images taken from below of the double oscillating jumps at different times. $T_1=0.28$ s is the experimentally measured period, which is very close to theoretical prediction (0.27 s) for $n=1$ mode. Right: representation of surface deformation for $n=1$ mode -for the same corresponding relative time. See movie M2 in the Supplemental Material \cite{SuppMat} for an animated representation of the dual jet oscillation.}
    \label{fig:double_jets}
\end{figure}

According to the theory, the first anticipated mode that is off center is the mode with $n=1$, associated with the Bessel function of the first kind $J_1$. As the first antinode is located at a distance $r_1=0.34 \, R_\text{sub}$ from the disk center and the second antinode is at the disk's border, i.e., $(n,l) = (1,2)$, this mode can therefore be excited by placing the jet at $r_1$. The predicted period of oscillation for this mode is $0.72 \times T_0$ where $T_0$ is the period of the $(0,1)$ mode for a given disk radius. These predictions are consistent with the experimental findings depicted in Fig. \ref{fig:higher_modes} when the distance of the jet to the center $r$ approaches $0.34 \, R_\text{sub}$. If we further displace the antinode, the next expected mode is the mode $(n,l) = (2,2)$ which corresponds to the Bessel function of the first kind $J_2$. In this case, the mode is expected to occur when the jet is positioned at a distance $r_2=0.45 \, R_\text{sub}$ from the disk center. The predicted period of oscillation is $0.56 \times T_0$. Once again, the experimental data align remarkably well with these predictions. The spatial representation of the amplitude of the modes $(1,2)$ and $(2,2)$ are illustrated as insets in Fig. \ref{fig:higher_modes}. 

Interestingly, according to the model, the $(n,l) = (1,2)$ mode displays two symmetric maxima for $r = 0.34 \, R_\text{sub}$ located on either side of the disk center. These maxima oscillate synchronously but with opposite phase as depicted on the spatiotemporal representation (right side) of Fig. \ref{fig:double_jets}. Motivated by this observation, we investigated experimentally the possibility of synchronizing the oscillations of two hydraulic jumps by placing them at the antinodes of the $(n,l) = (1,2)$ mode. The experiments were performed with a disk radius of $4.5$ cm. We indeed observed two hydraulic jumps oscillating synchronously with the period of the  $(n,l) = (1,2)$ mode but in phase opposition.

Finally, by combining our observations with previous works in the literature, we can propose a tentative explanation for the origin of the coupling between the hydraulic jump and surface waves. This explanation can be summarized in 3 steps. First, an hydraulic jump occurs due to the interplay between two phenomena: the flow's inertia, which tends to open the jump, and the hydrostatic pressure created by the outer water layer, which tends to close the jump. As a result, the radius of the jump decreases as the flow rate $Q$ decreases \cite{taniWaterJumpBoundary1949} and the outer water layer height $H$ increases \cite{ellegaardExperimentalResultsFlow1996}, eventually closing beyond certain threshold values of these two parameters. 
Second, the phase diagram shown in Fig. \ref{fig:diagramme} indicates that the oscillation instability occurs in the intermediate flow rate region between the stationary hydraulic jump and the closed state. In this intermediate region, small perturbations can trigger the transition between the closed and open states.
Lastly, surface gravity waves induce perturbations in the height of the outer water layer (for example due to cavity modes), which can, consequently, initiate the oscillations of the hydraulic jump. Conversely, the oscillations of the hydraulic jump generate surface gravity waves, which are amplified when they match with one of the disk cavity mode, thus providing a feedback mechanism for modes selection and amplification. To validate this hypothesis, we observed the variations in the height of the water layer around the jump (see movie M3 in the Supplemental Material \cite{SuppMat}) using the Schlieren visualization method \textcolor{red}{ \cite{moisySyntheticSchlierenMethod2009,wildemanRealtimeQuantitativeSchlieren2018}}. These observations reveal variations in the thickness of the water layer, amounting to approximately 25\% of the mean height value (3.9 mm), induced by surface waves. Moreover, we measured and plotted the evolution of the height $H$ of the outer water layer and compared it to the evolution of the radius of the jump (see Fig. SM1 in Supplemental Material \cite{SuppMat} in Supplementary Materials). This figure clearly demonstrate the synchronization between the two parameters' evolution, with a slight delay of approximately 10~ms providing evidence for the causality of the height variations over radius variation. Additionally, the figure clearly shows that the hydraulic jump opens below a critical height value as expected from data in the literature.

In conclusion, we report for the first time the stable periodic oscillations of a circular hydraulic jump. Our findings demonstrate that the oscillations originate from the coupling between surface waves eigenmodes and the hydraulic jump. Furthermore, we demonstrate that two hydraulic jumps can oscillate synchronously (in phase opposition) when placed at specific locations. This discovery paves the way for further exploration of the complex interaction between multiple oscillating jumps and surface waves. \\

The authors acknowledge the support of the French Agence Nationale de la Recherche (ANR), under Grant No. ANR-21-CE30-0014 (project IJET) and thank T. Bohr and A. Andersen for fruitful discussions and preliminary experiments, as well as Loïc Lam for his technical assistance.

\providecommand{\noopsort}[1]{}\providecommand{\singleletter}[1]{#1}%

\end{document}